\documentclass[
reprint,
showpacs,
showkeys,
 amsmath,amssymb,
 aip,
jcp,
letter
]{revtex4-1}

\usepackage{graphicx} % Include figure files
\usepackage{dcolumn} % Align table columns on decimal point
\usepackage{bm} % bold math
\usepackage[page]{appendix}
\usepackage{amsmath}
\usepackage{float}
\usepackage{scalerel}
\usepackage{subfigure}
\usepackage{color}
\usepackage[hidelinks]{hyperref} % add hypertext capabilities
\begin{document}

\title{Segmental motion and dynamic structure factor of discrete Gaussian semiflexible chains with arbitrary stiffness along the contour}
\title{Detailed dynamics of discrete Gaussian semiflexible chains with arbitrary stiffness along the contour}
%\title{Dynamics of Gaussian chains with variable stiffness: The whip polymer.}

\author{Andr\'{e}s R. Tejedor}
\affiliation{Department of Chemical Engineering, Universidad Polit\'{e}cnica de Madrid, Jos\'{e} Guti\'{e}rrez Abascal 2, 28006, Madrid, Spain}
\author{Jaime R. Tejedor}
\affiliation{Faculty of Physics, Department of Theoretical Physics, Universidad Complutense de Madrid, Plaza de las Ciencias, Ciudad Universitaria, Madrid 28040, Madrid, Spain}
\author{Jorge Ram\'{i}rez}
\email{jorge.ramirez@upm.es}
\affiliation{Department of Chemical Engineering, Universidad Polit\'{e}cnica de Madrid, Jos\'{e} Guti\'{e}rrez Abascal 2, 28006, Madrid, Spain}

\date{\today}

\begin{abstract}

We revisit a model of semiflexible Gaussian chains proposed by Winkler \textit{et al}, solve the dynamics of the discrete description of the model and derive exact algebraic expressions for some of the most relevant dynamical observables, such as the mean-square displacement of individual monomers, the dynamic structure factor, the end-to-end vector relaxation and the shear stress relaxation modulus. The mathematical expressions are verified by comparing them with results from Brownian dynamics simulations, reporting an excellent agreement. Then, we generalize the model to linear polymer chains with arbitrary stiffness. In particular, we focus on the case of a linear polymer with stiffness that changes linearly from one end of the chain to the other, and study the same dynamical functions previously presented. We discuss different approaches to check whether a polymer has constant or heterogeneous stiffness along its contour. Overall, this work presents a new insight for a well known model for semiflexible chains and provides tools that can be exploited to compare the predictions of the model with simulations of coarse grained semiflexible polymers.

%The dynamics of discrete semiflexible chains with Gaussian segments in diluted conditions is studied here. We start from a model proposed by Winkler \textit{et al} to study polymers with bending rigidity. First, we study the dynamics of homogeneous semiflexible chains, calculating exact expressions for the mean-square displacement, the dynamic structure factor, the end-to-end vector relaxation and the shear stress relaxation function. The results are verified by means of Brownian dynamics simulations, reporting an excellent agreement between both approaches. Second, we generalize the model to linear polymer chains with arbitrary stiffness and we also demonstrate that the methodology is general enough so it can be used in a wide range of systems. Particularly, we focus on the case of a linear polymer with stiffness that changes linearly from one end of the chain to the other. We study the same dynamical functions presented before and we also compare with simulations, confirming that the physics of this system is well described by the model. This work presents a new insight for the studied model that can be exploited by means of theory and simulations of polymers.
\end{abstract}

%\keywords{Semiflexible linear polymers, Brownian dynamics simulations, }
\maketitle

%\tableofcontents

\section{\label{sec:intro}Introduction}

The bending rigidity is a crucial feature exhibited by many polymers of biological interest such as double-stranded DNA or actin\cite{bustamante1994,kas1996,ober2000} that critically determines the functionality of those biomolecules. More precisely, it has been shown that the semiflexible conformation along the backbone of biopolymers is fundamental in several processes such as protein assembly\cite{garaizar2020}, dynamical behavior and stretching of DNA\cite{marko1995,shusterman2004}, enzymatic catalysis\cite{richard2019} or organisation of active filaments\cite{vliegenthart2020,claessens2006,winkler2020}. Thus, the understanding of how molecular stiffness affects the structural and dynamical properties of macromolecules is crucial both in biology and polymer science.

%Many macromolecules of biological importance such as double-stranded DNA (dsDNA), or proteins (e.g. actin, tubulin)  stiffness along their backbone\cite{bustamante1994,kas1996,ober2000}, and this stiffness affects their structural and dynamical properties which, in turn, have an impact in their biological functionality. In fact, it has been shown that the semiflexible conformation of some biomolecules is fundamental in several processes such as protein assembly\cite{garaizar2020}, dynamical behavior and stretching of DNA\cite{marko1995,shusterman2004}, enzymatic catalysis\cite{richard2019} or organisation of active filaments\cite{vliegenthart2020,claessens2006,winkler2020}. Thus, the understanding of how molecular stiffness affects the structural and dynamical properties of macromolecules is crucial both in biology and polymer science. 

Over the last decades, simulation of semiflexible molecules has emerged as a very useful tool to strengthen the understanding of the behavior of such systems. The most frequently used model to study semiflexible chains is the worm-like chain (WLC)\cite{kratky1949}, a non-linear model which describes the polymer conformation as an inextensible and differentiable curve with some bending elasticity. Initially proposed by Kratky and Porod, the WLC model has been extensively employed since its publication\cite{marko1995,bathe2008,wilhelm1996,bullerjahn2011,koslover2014,hinczewski2009,marantan2018}, and it has been successfully used to mimic  the dynamics of different biomolecules such as dsDNA\cite{marko1995,Peters_2010} or microtubules\cite{bathe2008}. For typical stiff macromolecules, the stretching modulus is much larger than the bending modulus, so that the chains are almost inextensible at length scales smaller than the persistence length, which is a crucial characteristic of the WLC. However, due to the non-linearity of the WLC model, the analytical treatment of the observables becomes non-trivial. Only a few of the conformational properties can be exactly calculated, and most dynamical properties are only accessible by means of simulations or after introducing approximations\cite{Kleinert_2006,Hsu2010,marantan2018}. 

Several alternative models for semiflexible chains have been proposed from different theoretical approaches. Harris and Hearst presented an analytically tractable model that introduces the stiffness in a similar way to the WLC model but restricting the average contour length instead of the bond length\cite{Harris_1966, Hearst_1967}. Bixon and Zwanzig\cite{Bixon_1978} studied the dynamics of linear semiflexible chains with constant bond length by applying linear response methods.
Other approaches, inspired by the WLC model, have employed a path integral formulation of the distribution function \cite{Freed_1972,Kleinert_2006} which formally permits to calculate most conformational statistics at equilibrium but they are difficult to use when the aim is to study the dynamical response. 
Barkema \textit{et al.} \cite{barkema2012,barkema2014} proposed a non-linear bead-spring model of semiflexible extensible chains which, after some approximations, can be used to predict several dynamical observables. In fact, Barkema's model recovers the WLC dynamics in the limit of large values of the Lagrangian multipliers used in the stretching and bending terms of the Hamiltonian. This approach has been exploited in simulations\cite{barkema2014,panja2015,barkema2017} and in calculations of dynamical functions such as the mean-squared displacement or the end-to-end relaxation\cite{barkema2014}, but most of the analytical expressions for the observables can only be obtained after linearization of the dynamics.

One model that has gained much attention in the last two decades is the one proposed by Winkler \textit{et al.}\cite{winkler1994,harnau1995} as a continuation of the work of of Bixon and Zwanzig\cite{Bixon_1978}. In Winkler's innovative studies, they applied the maximum entropy principle to develop a fully linear bead-spring model for semiflexible chains. Stiffness is introduced in a similar way to the WLC model, but instead of restricting the extensibility of each bond, only the mean-square bond length is constrained. Therefore, it can be regarded as an extension of the Rouse model, adding stiffness but preserving the Gaussian statistics for each chain segment. The main advantage of this model consists in formulating its Hamiltonian as a bilinear form and thus the partition function can be exactly calculated. However, even in the limit of very high stiffness, due to the Gaussian character, the contour length of the chain is allowed to fluctuate and the WLC and Winkler's model behave differently. Winkler's model has been used extensively, usually in the continuous version and including hydrodynamic interactions\cite{harnau1996,winkler2003,petrov2006}, with shear flow\cite{winkler2010} and with activity\cite{martin2019,philipps2022}. Additionally, the work has inspired similar discrete approaches to study more complex polymeric architectures, like stars, dendrimers\cite{Dolgushev_2009}, tree-like structures\cite{Dolgushev_2009a}, hyperbranched polymers\cite{Furstenberg_2013} and rings\cite{Dolgushev_2011,philipps2022}, and even chains and dendrimers with heterogeneous semiflexibility\cite{Dolgushev_2010}. 

%Winkler \textit{et al.}\cite{winkler1994,harnau1995} proposed a fully linear bead-spring model for semiflexible chains. Semiflexibility is introduced in a similar way to the WLC model. However, instead of restricting the extensibility of each bond, only the mean-square bond length is constrained, and thus, it can be regarded as an extension of the Rouse model, adding stiffness but preserving the Gaussian statistics for each chain segment. Due to the linear character of the model, the partition function can be exactly calculated. However, even in the limit of very high stiffness, due to the Gaussian character, the contour length of the chain is allowed to fluctuate and the WLC and Winkler's model behave differently. Winkler's model has been used extensively, usually in the continuous version and including hydrodynamic interactions\cite{harnau1996,winkler2003,petrov2006}, with shear flow\cite{winkler2010} and with activity\cite{martin2019}. A similar discrete approach, inspired by the work of Bixon and Zwanzig\cite{Bixon_1978} has been used to study more complex polymeric architectures, like stars, dendrimers\cite{Dolgushev_2009}, tree-like structures\cite{Dolgushev_2009a}, hyperbranched polymers\cite{Furstenberg_2013} and rings\cite{Dolgushev_2011}, and even chains and dendrimers with heterogeneous semiflexibility\cite{Dolgushev_2010}. 

For fully flexible polymer chains, the Rouse model is the touchstone upon which most theories, such as the tube model for entangled polymers, are formulated, because it captures connectivity, flexibility and local dissipation, as well as because all observables can be expressed analytically, both in the continuous form  \cite{doi1988theory} and in the discrete form\cite{Likhtman_2012}. We believe that Winkler's model can play a similar role as the Rouse model but for semiflexible polymers. Additionally, Winkler's model can be the ideal choice when studying coarse-grained semiflexible chains, in which each bead typically represents many monomer units, the stretching modulus may not be negligible with respect to the bending modulus. 

Although Winkler formulated his model in terms of discrete beads and bonds, most of the works use a continuous version of it \cite{winkler2003,winkler2010}, which is a reasonable approach for chains with a high number of beads. Nevertheless, in many practical applications, it is more useful to deal with semiflexible bead-spring chains of small and moderate molecular weight. For example, Brownian Dynamics simulations are a very extended tool to study the dynamics of polymers, both flexible and stiff. Having a fully exact approach to generate equilibrium semiflexible chains and a reference point against which the results of simulations can be compared would be very useful.

In this work, we revisit the model of semiflexible polymers proposed by Winkler (Section \ref{sec:semiflexible}), and propose a fully analytical treatment to extract the most  relevant dynamical observables in the discrete description. In particular, we provide exact expressions for the monomeric mean-squared displacement (Section \ref{subsec: MSD_semiflexible}), end-to-end vector relaxation (Section \ref{subsec: ete_Semiflexible}), shear-stress relaxation modulus (Section \ref{subsec: Gt_semiflexible}) and dynamic structure factor (Section \ref{subsec: DSF_semiflexible}). A similar approach, but for a smaller set of observables (dynamic modulus and dielectric relaxation), has been proposed previously\cite{Dolgushev_2009,Dolgushev_2010}. Our analytical predictions are in very good agreement with results from Brownian dynamics simulations of the same model. Additionally, our analytical approach provides a recipe to generate equilibrium ensembles of semiflexible chains easily that can be used to start simulations that follows the same equations of motion. Additionally, the model can be generalized for chains of arbitrary bond length, stiffness and even longer distance correlations between bonds along the contour, as long as the model remains linear. Particularly, we explore the dynamics of linear polymer chains with linearly varying stiffness along the contour (Section \ref{sec:whip}). All the results in this work neglect hydrodynamic interactions, so it would be suitable for cases where these interactions are screened, such as in the case of melts of unentangled semiflexible polymers. 

\section{\label{sec:semiflexible}Semiflexible Gaussian polymer}

\begin{figure}[ht]
	\centering
	\includegraphics[width=\columnwidth]{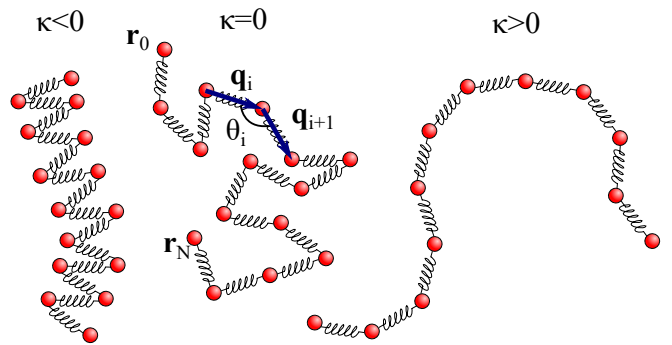}
	\caption{Semiflexible Gaussian chains with different values of the stiffness parameter $\kappa$: folded-chain ($\kappa<0$, left), Rouse chain ($\kappa=0$, center) and semiflexible polymer ($\kappa>0$, right).}
	\label{fig: Fig1}
\end{figure}

We use the model of semiflexible Gaussian chains introduced by Winkler \textit{et al.}\cite{winkler1994,harnau1995}. The advantage of this model is that the Hamiltonian can be expressed as a bilinear form in terms of the bead positions, which allows to exactly solve the dynamics of the chains by introducing the  proper normal coordinates. The linear semiflexible chain is represented by a series of $ N+1 $ beads with equal mass connected by $ N $ harmonic springs. We denote the position of each bead as $ \mathbf{r}_i $, with $ i=0,...,N $, and the connector vectors that links neighbouring beads as $ \mathbf{q}_i=\mathbf{r}_i-\mathbf{r}_{i-1} $, with $i=0,...,N$. The model presents the following constraints
\begin{equation}\label{eq: ConstrSemiflexible}
\begin{split}
	&\langle\mathbf{q}_i^2\rangle = b^2\qquad i = 1,...,N\\
	&\langle\mathbf{q}_i\mathbf{q}_{i+1}\rangle = \kappa b^2\qquad i = 1,...,N-1,
\end{split}
\end{equation}
where $ b^2 $ is the average bond length and $ \kappa = \langle \cos\theta_i\rangle$, where $\theta_i$ is the angle between consecutive connecting vectors $\mathbf{q}_{i}$ and $\mathbf{q}_{i+1}$. Thus, the parameter $\kappa$ controls the bending rigidity of the chain and can take values from -1 (completely folded chain) to 1 (rod limit). Here we restrict the value of $\kappa$ to the interval from $\kappa=0$, corresponding to a flexible Rouse chain, to $\kappa=1$, which is similar to a rigid rod, although the contour length can fluctuate. In Fig. \ref{fig: Fig1} we depict several examples of Gaussian chains with different  values of $\kappa$.

The potential energy of a semiflexible Gaussian chain in terms of the bond vectors can be written as:
\begin{equation}\label{eq: PotSemiflexibleq}
	U = \frac{3k_BT}{b^2}\left[\sum_{i=1}^N\lambda_i\mathbf{q}_i^2 - \sum_{i=1}^{N-1}\mu_i\mathbf{q}_i\mathbf{q}_{i+1}\right].
\end{equation}
The parameters $ \lambda_i,\,\mu_i $ are Lagrangian multipliers that can be easily computed by enforcing the constraints \eqref{eq: ConstrSemiflexible} and using the maximum entropy principle, as described in detail by Winkler\cite{winkler1994,winkler2003,Dolgushev_2009}:
\begin{equation}\label{eq: LagMulSemiflexible}
\begin{split}
\lambda_i=\frac{1}{2}\frac{1+\kappa^2}{1-\kappa^2},\quad i=2,...N-1\\
\mu_i =\frac{\kappa}{1-\kappa^2},\quad i=1,...N-1,
\end{split}
\end{equation}
and $\lambda_1=\lambda_N=1/2(1-\kappa^2)$. Here, the hydrodynamic interactions are neglected and thus the model may in principle be valid for melts of unentangled semiflexible linear chains, or blends of unentangled semiflexible chains diluted in a mesh of flexible molecules. 
The  potential energy in eq. \eqref{eq: PotSemiflexibleq} can be expressed as a bilinear form:
\begin{equation}\label{eq: PotentialQBQ}
	U = \frac{3k_BT}{b^2}\mathbf{Q}^TB\mathbf{Q},
\end{equation}
where $ \mathbf{Q}^T=(\mathbf{q}_1,...,\mathbf{q}_N) $, and the matrix $ B $ is a $N \times N$ constant, symmetric, positive semi-definite  and tridiagonal matrix with $ \lambda_i $ in the diagonal and $ -\mu_i/2 $ as off-diagonal elements. To study  observables such as the segmental motion or the dynamic structure factor, it is advantageous to rewrite the expression of the potential energy as a function of the bead positions by introducing a change of variables $ \mathbf{q}_i= \mathbf{r}_i -\mathbf{r}_{i-1}$:  
\begin{equation}\label{eq: PotentialRAR}
U = \frac{3k_BT}{b^2}\mathbf{R}^TA\mathbf{R},
\end{equation}
where we have defined $\mathbf{R}^T=(\mathbf{r}_0,...,\mathbf{r}_N)$. Here, $ A $ is a $(N+1) \times (N+1)$ constant, symmetric, positive semi-definite and pentadiagonal matrix (See SI).  

\subsection{Evolution of bead positions}

Given the potential energy \eqref{eq: PotentialRAR}, the overdamped evolution equation of the bead positions is given by
\begin{equation}\label{eq: Langevin}
	\zeta\frac{\partial \mathbf{r}_{n}}{\partial t}=-\frac{\partial U}{\partial \mathbf{r}_{n}}+\mathbf{f}_{n}(t),
\end{equation}
where $\zeta$ is the friction coefficient of one bead, $n$ ranges from 0 to $N$, and $\mathbf{f}_n(t)$ is a vector whose components are  Gaussian random variables with zero mean and variance determined by the fluctuation dissipation theorem\cite{van1992}:
\begin{equation}
	\left\langle f_{n\alpha} (t) f_{m\beta} (t')\right\rangle=2k_BT\zeta\delta_{nm}\delta_{\alpha\beta}\delta(t-t').
\end{equation}
where $k_B$ is the Boltzmann's constant, $T$ is the temperature, $n$ and $m$ represent different bead indices and $\alpha,\beta=x, y, z$. Due to the linear nature of the model, the $x, y$ and $z$ coordinates of the chain beads evolve independently and thus, for the sake of simplicity,  in the remaining of the paper we will show the solution to the dynamics of the chains only along one of the Cartesian coordinates. Therefore, differenciating the potential with respect to $ \mathbf{r}_i $, the evolution equation can be rewritten as
%\begin{equation}\label{eq: LangevinMatrix}
%	\zeta d\mathbf{R}_{\alpha}=-kA\mathbf{R}_{\alpha}dt+d\mathbf{W}_\alpha,
%\end{equation}
\begin{equation}\label{eq: LangevinMatrix}
	\zeta\frac {\partial \mathbf{R}_{\alpha}}{\partial t}=-kA\mathbf{R}_{\alpha} +\mathbf{F}_{\alpha}(t),
\end{equation}
 where $ k=3k_BT/b^2$, and the vector $\mathbf{F}_\alpha(t)$ contains the stochastic variables $f_{n\alpha} (t)$. The explicit solution to equation Eq. \eqref{eq: LangevinMatrix} is given by:
 \begin{equation}\label{eq: LangevinSolution}
 	\mathbf{R}_\alpha(t)=e^{-\frac{At}{\tau}}\mathbf{R}_\alpha(0)+\frac{1}{\zeta}\int^t_0dt'e^{-\frac{A(t-t')}{\tau}}\mathbf{F}_\alpha(t'),
 \end{equation}
 where we have defined the relaxation time $\tau=\zeta b^2/3k_BT$. Using the explicit solution of eq. \eqref{eq: LangevinSolution}, exact analytical expressions can be derived for different dynamical observables. In the solution we use $ \tau $ and $ b $ as the units of time and length, respectively. Alternatively, we can discretize the equations of motion \eqref{eq: Langevin}, run Brownian dynamics (BD) simulations using different values of the stiffness parameter, and analyze the trajectories to extract the desired observables. %In the following sections, we infer the analytical expressions of several observables, compare them with the results from BD simulations, and discuss the effect of the rigidity parameter $\kappa$ (See SI for simulation details.).
 
 \subsection{Segmental motion}\label{subsec: MSD_semiflexible}

Using the matrix notation introduced above, a general form of the mean-squared displacement (MSD) can be expressed as a matrix of $ (N+1)\times (N+1) $ elements where the element ($ i,j $) gives the cross-correlation between the displacement of the $ i $th and $ j $th beads:
\begin{equation}\label{eq: MSDef}
	\mathbf{g}(t)=3\left\langle \left(\mathbf{R}_\alpha(t)-\mathbf{R}_\alpha(0)\right)\left(\mathbf{R}_\alpha(t)-\mathbf{R}_\alpha(0)\right)^T \right\rangle,
\end{equation}
where the factor 3 appears due to the three independent spatial coordinates ($ \alpha=x,y,z $). 
%Since the system is at equilibrium and is isotropic, and all three spatial directions are equivalent, we only compute the solution along one of them. 
The MSD of individual beads is given by the diagonal elements of the matrix $\mathbf{g}(t)$, \emph{i.e.} $ g_1(i,t)=\mathbf{g}_{ii}(t) $. 

Substituting the solution of $\mathbf{R}_\alpha$ from Eq. \eqref{eq: LangevinSolution} in Eq. \eqref{eq: MSDef}, expanding and simplifying several null terms (see SI), we find:
\begin{equation}\label{eq: MSDmsol}
\begin{split}
\mathbf{g}(t) =& 3\left(e^{-\frac{At}{\tau}} - \mathbb{I}\right)\left\langle\mathbf{R}_\alpha\mathbf{R}_\alpha^T\right\rangle_0\left(e^{-\frac{At}{\tau}} - \mathbb{I}\right)\\  
+ & \frac{6k_BT}{\zeta}\int^t_0dt'\,e^{\frac{-2A(t-t')}{\tau}}.
\end{split}
\end{equation}
In the last expression, the average of position vectors $\mathbf{R}_\alpha$ is evaluated at time zero (\textit{i.e.}, at equilibrium). The expression can be further simplified by using $\tau$  and $b$ as units of time and length, respectively, to get 
\begin{equation}\label{eq: MSDmNDsol}
\begin{split}
	\mathbf{g}(t) & = 3\left(e^{-At} - \mathbb{I}\right)\left\langle\mathbf{R}_\alpha\mathbf{R}_\alpha^T\right\rangle_0\left(e^{-At} - \mathbb{I}\right) \\
	& + 2\int^t_0dt'\,e^{-2A(t-t')}.
\end{split}
\end{equation}
Please, note that, for the sake of simplicity, we will use the same notation to refer to non-dimensional variables (such as $\mathbf{g}$, $\mathbf{R}_\alpha$ or $t$) and their dimensional counterparts. The latter expression for the MSD matrix can be simplified by introducing the diagonalization of $A$ or, equivalently, by using normal coordinates. In contrast to fully flexible (Rouse) case, the analytical spectral decomposition of $A$ for semiflexible chains is not possible and, thus, the matrix $A$ must be diagonalized numerically. Substituting the spectral decomposition $ A=P^TDP $ in Eq. \eqref{eq: MSDmNDsol} we obtain
%\begin{equation}\label{eq: MSDmwithPDP}
%\begin{split}
%\mathbf{g}(t)&  = \\
%& 3P^T\left(e^{-Dt} - \mathbb{I}\right)P \left\langle\mathbf{R}_\alpha\mathbf{R}_\alpha^T\right\rangle_0  P^T\left(e^{-Dt} - \mathbb{I}\right)P \\
%& + 2 \int^t_0dt'\,P^Te^{-2D(t-t')}P.
%\end{split}
%\end{equation}
%The integral of the diagonal matrix in the last term can be easily evaluated and expressed as a function of the eigenvalues $ \{\nu_0,...,\nu_N\} $ of the matrix $A$. Note that the eigenvalue $ \nu_0=0 $ is associated with the difussive motion of the center of mass. The MSD matrix is thus reduced to:
\begin{equation}\label{eq: MSDmFINsol}
\begin{split}
	\mathbf{g}(t) &= \\
	& 3P^T\left(e^{-Dt} - \mathbb{I}\right)P \left\langle\mathbf{R}_\alpha\mathbf{R}_\alpha^T\right\rangle_0 P^T\left(e^{-Dt} - \mathbb{I}\right)P \\
	& + 2P^TD^*P.
\end{split}
\end{equation}where
\begin{equation}\label{eq: IntDstar}
D^* = \begin{pmatrix}
t & 0 & \cdots & 0\\
0 & \frac{1-e^{-2\nu_1t}}{2\nu_1} & \ddots & \vdots\\
\vdots & \ddots & \ddots & 0\\
0 & \cdots & 0 & \frac{1-e^{-2\nu_Nt}}{2\nu_N}\\
\end{pmatrix},
\end{equation}
and $ \{\nu_0,...,\nu_N\} $ represent the eigenvalues of $A$. In Eq. \eqref{eq: MSDmFINsol}, the average $ \left\langle\mathbf{R}_\alpha\mathbf{R}_\alpha^T\right\rangle_0 $ must be evaluated at equilibrium, and it represents a matrix whose element $(n,m)$ is given by the scalar product $\left\langle\mathbf{r}_n\textbf{r}_m\right\rangle_0$. Without loss of generality, we can arbitrarily set the position of the first bead of all equilibrium chains at the origin of coordinates to facilitate the calculation. It is well-known that in the flexible limit (Rouse chain) the average yields $\left\langle\mathbf{r}_n\textbf{r}_m\right\rangle_0=\min{(n,m)}$ in non-dimensional units, whereas in the fully rigid limit $ (\kappa\rightarrow 1) $ it gives $\left\langle\mathbf{r}_n\textbf{r}_m\right\rangle_0=n\cdot m$. In the general semiflexible case, we can use the scalar product of the bond vectors $\mathbf{q}_n\cdot\mathbf{q}_m$ which, similarly to the freely rotating chain\cite{doi1988theory}, can be expressed as a function of the average cosine of the angle between consecutive bonds:
\begin{equation}\label{eq: Productq}
\left\langle\mathbf{q}_n\mathbf{q}_m\right\rangle_0=\kappa^{|n-m|}.
\end{equation} \label{eq: ProductRR}
Using $\mathbf{r}_n=\mathbf{q}_n+\mathbf{r}_{n-1}$, we can write the scalar product $ \left\langle\mathbf{r}_n\textbf{r}_m\right\rangle_0 $ as a sum of products $ \left\langle\mathbf{q}_n\textbf{q}_m\right\rangle_0 $ and, using Eq. \eqref{eq: Productq}, the scalar product can be asserted as (See SI for details):
\begin{equation}\label{eq: ProductSemiflexible}
\left\langle\mathbf{r}_n\mathbf{r}_m\right\rangle_0=\sum_{k=1}^{n}\sum_{l=1}^{m}\kappa^{|k-l|}.
\end{equation}
In the previous expression, it is easy to verify that the Rouse and rigid rod limits are recovered when the corresponding values of $\kappa$ are used. In summary, the segmental MSD matrix $ \mathbf{g}(t) $ in Eq. \eqref{eq: MSDmFINsol} can be computed exactly using the scalar product of bead positions in Eq. \eqref{eq: ProductSemiflexible}. Only the diagonalization of the constant matrix $ A $ requires numerical analysis. 
%The MSD function $ g_1(i,t) $ is simply extracted as the diagonal of $ \mathbf{g}(t) $.

\begin{figure}[ht]
	\centering
	\includegraphics[width=\columnwidth]{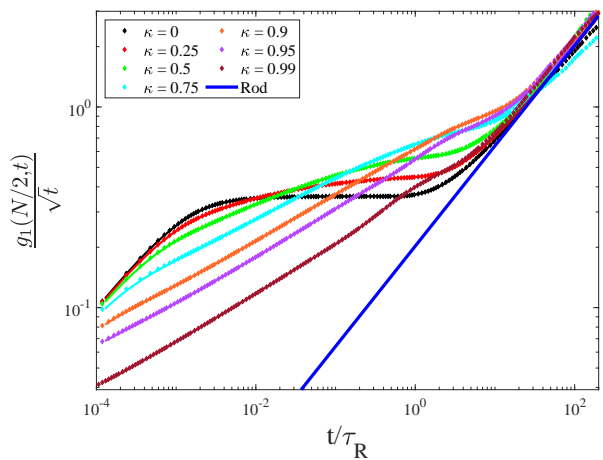}
	\caption{Mean-squared displacement (MSD) of the middle bead of semiflexible Gaussian chains with semiflexibility parameter $ \kappa=0.0,\,0.25,\,0.5,\,0.75,\,0.9,\,0.95,\,0.99$, from BD simulations (symbols) and from Eq. \eqref{eq: MSDmFINsol} (solid lines). MSD data are divided by $\sqrt{t} $ and time is divided by $\tau_R $. The solid blue line indicates the MSD of the central bead of a rigid rod, which corresponds to the the diffusion of the center of mass of all chains, regardless of their rigidity.}
	\label{fig: Fig2}
\end{figure}

In Fig. \ref{fig: Fig2}, the MSD of the central bead $ g_1(N/2,t)=\mathbf{g}_{N/2,N/2}(t) $ of a chain with $ N=50 $ springs from BD simulations (symbols) and from Eq. \eqref{eq: MSDmFINsol} are shown. The data from simulations (the MSD as well as all other observables in the remaining of this article) have been obtained by using an efficient correlator technique \cite{Ramirez_2010}. The $g_1(N/2,t)$ axis has been divided by $\sqrt{t}$ in order to highlight the characteristic subdiffusive regime of the Rouse model ($t^{1/2}$), and the time axis has been divided by $ \tau_R \propto N^2$, the terminal time of the purely flexible chain.
The rod-like limit ($\kappa=1$) cannot be resolved analytically or by BD simulations for this model, but we can use the fact that the central bead of a rigid rod is equal to the center of mass of the chain and that the diffusion of the center of mass is independent of the value of $\kappa$, in the free draining, dilute case. Therefore, the MSD of the central bead of a rod must be equal to that of the center of mass of a Rouse chain, \emph{i.e.} $g_3(t)= 6D_Gt$, with $ D_G=k_BT/N\zeta $. Please note that in the rod-like limit of Winkler's model, the position of the central bead fluctuates around the center of mass of the chain and, thus, the mean-squared displacement of the middle bead should be larger than that of a rigid rod at short and intermediate times.

The results from analytical calculations and BD simulations show an excellent agreement, which supports the accuracy of the analytical expression \eqref{eq: MSDmFINsol} as well as of the BD algorithm. 
%except at very early time, where the data slightly disagree due to the precision of the simulation algorithm ($ \propto \sqrt{dt} $). 
The fully flexible chain (black data, $ \kappa=0 $) shows the expected subdiffusive regime ($t^{1/2}$) when $\tau_0 \ll t \ll \tau_R$, where $\tau_0$ is the relaxation time of one spring, before reaching the terminal Fickian diffusion at times much longer than the Rouse time $\tau_R$, when the diffusion of the central monomer is dragged by the diffusion of the center of mass. As mentioned above, the MSD of the center of mass is the same for all chains at very long times, regardless of their stiffness. 

As the chain becomes stiffer, two effects arise: First, the subdiffusive regime starts at shorter times than for the Rouse chain. By increasing the stiffness, the shortest relaxation time gets increasingly smaller since the lateral displacement of beads in the perpendicular direction to the end-to-end vector becomes more restricted, which represents an additional constraint on the harmonic interaction between adjacent beads. Second, the power law at intermediate times becomes larger than 1/2. The approximate power law $ g_1(N/2,t)\propto t^{0.7} $ for a stiffness parameter $\kappa \gtrsim 0.75$ is in agreement with the results obtained by Barkema \textit{et al.}\cite{barkema2014}. This higher slope is necessary since the beads depart from a slower MSD and need to reach the same Fickian diffusion regime after the terminal time. It is interesting to note that, as the stiffness parameter $\kappa$ increases, there is a region closer to the terminal time where the central bead moves faster than it does in a Rouse chain (see for instance the curve for $\kappa=0.9$ near $t/\tau_R=1$). The width of this region decreases as $\kappa$ grows, and the region finally disappears for very stiff chains ($\kappa \rightarrow 1$). This is due to the anisotropy of the MSD of beads. The bending interaction restricts their lateral motion, but the beads can slide in the direction of the end-to-end vector more easily. %Finally, it is worth highlighting that the stiffness parameter must be greater than 0.5 to have a considerable impact on the MSD of the middle bead, and the difference between $ \kappa =0.9 $ and $ \kappa=0.95 $ is greater than the difference between the flexible chain and the polymer with $ \kappa=0.25 $.

Finally, the stiffer the chain the closer the MSD of the middle monomer becomes to the rigid rod limit, given by the blue line in Fig. \ref{fig: Fig2}. It is important to remind that, although a $\kappa\to1$ parameter guarantees a rod-like configuration of the chain, the mobility of the individual beads is larger than in a pure rod. Even though the configuration of the chain is a straight line, the beads still have some freedom to move in the longitudinal direction, due to the Gaussian character of the bonds and the fluctuations of the molecular length. A pure rod, such as the one considered in the Doi-Edwards book\cite{doi1988theory}, has a constant length. This is why the MSD of the central monomer of the stiffest chain considered in this work (see Fig. \ref{fig: Fig2}) is significantly faster than the theoretical rod limit at short and intermediate times. 
%As the rigidity parameter $\kappa$ increases, the MSD of the central monomer gets closer and closer to the theoretical limit of the rigid rod, given by the blue line in Fig. \ref{fig: Fig2}. It is important to remind that, although a $\kappa\to1$ parameter guarantees a rod-like configuration of the chain, the mobility of the individual beads is larger than in a pure rod. Even though the configuration of the chain is a straight line, the beads still have some freedom to move in the longitudinal direction, due to the Gaussian character of the bonds and the fluctuations of the molecular length. A pure rod, such as the one considered in the Doi-Edwards book\cite{doi1988theory}, has a constant length. This is why the MSD of the central monomer of the stiffest chain considered in this work (see Fig. \ref{fig: Fig2}) is significantly faster than the theoretical rod limit at short and intermediate times. 

%Another interesting aspect that can be barely appreciated in Fig. \ref{fig: Fig2} is that, as the chains get stiffer, the relaxation time becomes longer. This detail can be better inspected with the autocorrelation of the end-to-end vector, which we explore in a later section.

\subsection{Dynamic structure factor}\label{subsec: DSF_semiflexible}

%Despite the striking differences in the segmental motion of chains with different stiffness, the relaxation times barely seems to increase with the rigidity, although it should heavily depend on the conformation. 
Besides the MSD, additional information about the inter-particle correlations and their evolution with time can be inspected with the dynamic structure factor $S(\mathbf{q},t)$. This observable was not studied in the works of the group of Blumen\cite{Dolgushev_2009,Furstenberg_2012}, where they studied a similar semiflexible chain. In the work of Harnau \textit{et al.} on the continuous model of semiflexible polymers\cite{harnau1996,harnau1995}, they provide a solution for the dynamic structure factor discussing the effect of stiffness for different regimes. Here, we present the solution for the discrete model neglecting hydrodynamic interactions so that this work provides a new insight in contrast to the work of Harnau \textit{et al.}\cite{harnau1995,harnau1996}. The coherent single chain dynamic structure factor $S(\mathbf{q},t)$ 
%of this discrete semiflexible Gaussian model 
gives information about the dynamics of polymer chains at different length scales depending on the magnitude of the scattering vector $\mathbf{q}$. Experimentally $S(\mathbf{q},t)$ can be accessible from neutron spin-echo experiments on samples that contain a small fraction of protonated chains in a mesh of deuterated polymers. The coherent single chain dynamic structure factor is defined as:
%One observable that was not considered in the original work of the group of Blumen\cite{Dolgushev_2009,Furstenberg_2012}, was the dynamic structure factor $S(\mathbf{q},t)$. In the work of Harnau \textit{et al.} on the continuous model of semiflexible polymers\cite{harnau1996,harnau1995}, they provide a solution for the dynamic structure factor discussing the effect of stiffness for different regimes. Here, we present the solution for the discrete model neglecting hydrodynamics interactions so that this work provides a new insight in contrast to the work of Harnau \textit{et al.}\cite{harnau1995,harnau1996}. The coherent single chain dynamic structure factor $S(\mathbf{q},t)$ of this discrete semiflexible Gaussian model, which provides information about the dynamics of the polymer chain at different length scales depending on the magnitude of the scattering vector $\mathbf{q}$, can be easily accessible from neutron spin-echo experiments on samples that contain a small fraction of protonated chains in a mesh of deuterated polymers. The coherent single chain dynamic structure factor is defined as:
\begin{equation}\label{eq: DSFdef}
		S(\mathbf{q},t)=\frac{1}{N}\sum_{m,n}\langle\exp[i\mathbf{q}\cdot(\mathbf{r}_m(t)-\mathbf{r}_n(0))]\rangle,
\end{equation}
where $ i $ is the imaginary unit and the scattering vector $\mathbf{q}$ can be related to a characteristic correlation distance $d$ in non dimensional units through the expression $\mathbf{q}^2=1/d^2$. In the case of Gaussian chains, this expression can be simplified to\cite{doi1988theory}: 
\begin{equation}\label{eq: DSFdefSim}
	S(\mathbf{q},t)=\frac{1}{N}\sum_{m,n}\exp\left[-\frac{\mathbf{q}^2}{6}\phi_{mn}(t)\right],
\end{equation}
where 
\begin{equation}\label{eq: DSFPhi}
	\phi_{mn}(t)=3\left\langle(r_{m\alpha}(t)-r_{n\alpha}(0))^2\right\rangle,
\end{equation}
where the factor 3 appears again because the three spatial coordinates are equivalent and we calculate the solution only along one of them. The term $\phi_{mn}(t)$ can be regarded as the element $mn$ of a matrix $\mathbf{\phi} (t)$ which is very similar to the matrix $\mathbf{g}(t)$ that was used in the calculation of the segmental motion. However, in the present case, the element $\phi_{mn}(t)$ correlates the displacement of bead $m$ at time $t$ with respect to the position at time 0 of bead $n$, and the calculation is slightly more complicated. %In fact, only the diagonal terms of the matrix $\mathbf{\phi} (t)$ are exactly equal to the diagonal terms of the matrix $\mathbf{g}(t)$. 
For the calculation of $\mathbf{\phi} (t)$, we proceed in a very similar way as in the case of $\mathbf{g}(t)$
%, expanding the product in Eq. \eqref{eq: DSFPhi} and cancelling the cross terms 
(see SI for details), to get, in non-dimensional units:
\begin{equation}\label{eq: DSFsol}
\begin{split}
    \phi_{mn}(t)=&3\left(e^{-At}\mathbf{\rho} e^{-At}\right)_{mm}
		+3\mathbf{\rho}_{nn}\\
		&-3\left(e^{-At}\mathbf{\rho}\right)_{mn}-3\left(\mathbf{\rho} e^{-At}\right)_{nm}\\
		&+2\int_0^t\,\left(e^{-2A(t-t')}\right)_{mm}dt',
\end{split}
\end{equation}
where we have defined $\mathbf{\rho}=\langle\mathbf{R}_\alpha\mathbf{R}_\alpha^T\rangle_0$. Substituting Eq. \eqref{eq: DSFsol} into Eq. \eqref{eq: DSFdefSim}, the dynamic structure factor can be exactly computed, requiring only numerical analysis for the diagonalization of matrix $A$. 

\begin{figure*}[htb!]
	\centering
	\includegraphics[scale=0.35]{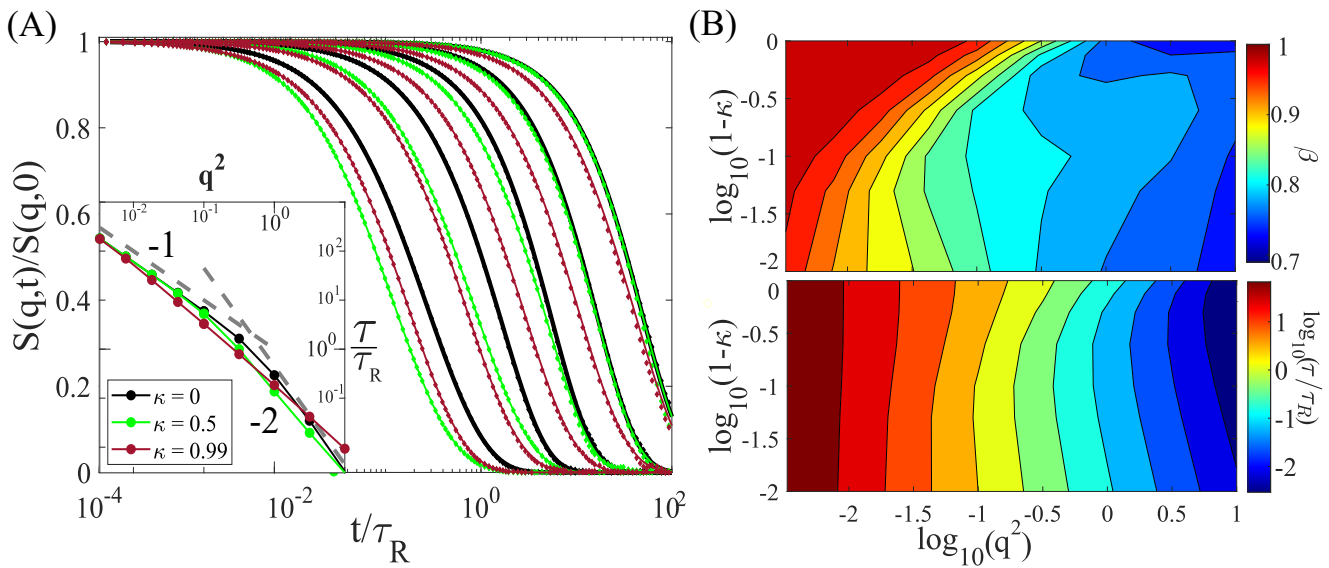}
	\caption{Dynamic structure factor of discrete semiflexible chains with different stiffness. (A) Dynamic structure factor from BD simulations (symbols) and analytical calculations (solid lines) for $ N=50 $ and stiffness $\kappa=0,\,0.5,\,0.99$. The scattering vectors $q$ are chosen so that the characteristic distance $d^2=1,3,9,27,81$, from left to right in the plot. Inset shows the relaxation time $\tau$ from the stretched exponential fit to $S(\mathbf{q},t)$ for three different values of $\kappa$ using a wider range of scattering vectors than in the main panel. Grey dashed lines show the power law of -1 and -2 as indicated.
	(B) Maps of the parameters $\beta$ (Top) and $\tau$ (Bottom) from the fits to a stretched exponential relaxation for all values of stiffness studied, using scattering vectors logarithmically spaced in the interval $\mathbf{q}^2=0.03,..,10$ in non-dimensional units. The stiffness is shown in the vertical axis as $1-\kappa$ in logarithmic scale to highlight the behavior near the rod-like regime.}
	\label{fig: Figure5}
\end{figure*}

In Fig. \ref{fig: Figure5} we show the results of the dynamic structure factor $S(\mathbf{q},t)$, from calculations (solid lines) and BD simulations (symbols), of chains with different stiffness, for a set of scattering vectors corresponding to distances $d^2=1,3,9,27,81$ that range from the bond distance to several times the end-to-end distance of a Rouse chain with $N=50$. 
The agreement between Eq. \eqref{eq: DSFsol} and the results from BD simulations is excellent for all values of $\kappa$ and $\mathbf{q}$, as can be seen in Fig. \ref{fig: Figure5}(A). For a fixed scattering vector, the dynamic structure factor for the Rouse chain decays more slowly than for stiffer chains. This is reasonable since, at fixed $\mathbf{q}$ (constant distance) the sampled scale is relatively larger for flexible chains (with smaller molecular size) than for semiflexible chains. As the chain becomes stiffer, the relaxation occurs much faster for a fixed $\mathbf{q}$, indicating a weaker correlation between bead positions at different times. At distances much longer than the end-to-end distance (very small $\mathbf{q}$), $S(\mathbf{q},t)$ is governed by the motion of the center of mass and thus all curves should converge. We can further explore the effect of stiffness by fitting the curves of $S(\mathbf{q},t)$ to a stretched exponential:
\begin{equation}\label{eq: StretchedExponential}
    f(t)=A\exp\left[-\left(\frac{t}{\tau}\right)^\beta\right],
\end{equation}
where we set the amplitude $A=1$, $\tau$ is the characteristic relaxation time, and $\beta$ is the stretching exponent with values between 0 and 1. For a discreet Rouse chain ($\kappa=0$) the dynamic structure factor is expected to decay as a single exponential at long distances\cite{doi1988theory} ($\mathbf{q}^2\ll 1$), with a relaxation time that scales as $\tau\propto\mathbf{q}^{-2}$. 
%Thus, the fit must result in stretching exponents close to one. 
On the other hand, at very short distances ($\mathbf{q}^2\gg 1$), the relaxation of $S(\mathbf{q},t)$ is not single exponential ($\beta<1$), and the characteristic time scales as $\tau\propto\mathbf{q}^{-4}$. The expected Rouse behavior is perfectly captured, as shown in the in the inset of \ref{fig: Figure5}(A) and in Figs. \ref{fig: Figure5}(B) ($\kappa=0$, upper region of the colour maps). Please bear in mind that the inset in Fig. \ref{fig: Figure5}(A) is plotted vs $\mathbf{q}^2$, so a decay with power law $-2$  implies $\tau\propto\mathbf{q}^{-4}$ as predicted for a Rouse chain\cite{doi1988theory}. As shown in Fig. \ref{fig: Figure5}(B), as the stiffness increases the value of $\beta$ decreases, even at relatively long distances, signalling the appearance of more complex dynamical behaviour than that of a Rouse chain. Still, for moderately stiff chains, the same scaling of the terminal time with the scattering vector as the Rouse chain is observed. 
Interestingly, the rod-like polymer exhibits roughly the same power law ($\tau\sim\mathbf{q}^{2.5}$) at all distances (see inset in Fig. \ref{fig: Figure5}(A)). 

It is important to note that the longest distance explored in Fig. \ref{fig: Figure5} ($d^2=81$) is still smaller than the end-to-end distance for moderately stiff chains with $\kappa\gtrsim0.7$ (see section \ref{subsec: ete_Semiflexible}) so that, unsurprisingly, the decay of the corresponding $S(\mathbf{q},t)$ occurs faster for stiff chains. 
In order to fairly compare the decay of $S(\mathbf{q},t)$ for chains with different semiflexibilities, we have rescaled the scattering vectors by the corresponding mean-square end-to-end distance of each semiflexible chain (see Fig. S1 in the Supplementary Info), observing that all the curves almost collapse, except for those that explore very small distances. The stretching exponents tend to 1 for all scattering vectors that explore distances beyond the end-to-end vector, and the relaxation time shows a simpler relationship with respect to $\mathbf{q}$.

\subsection{End-to-end vector relaxation}\label{subsec: ete_Semiflexible}
%Here we compute the end-to-end vector correlation function, so 
First, we express the end-to-end vector $\mathbf{R}_{ee,\alpha}(t)$ in terms of the bead positions as:
\begin{equation}\label{eq: Etebead}
\mathbf{R}_{ee,\alpha}(t)=\mathbf{v}^T\cdot \mathbf{R}_\alpha(t),
\end{equation}
where $\mathbf{v}^T=(-1,0,...,0,1)$ and, again, only one of the three equivalent spatial directions $\alpha=x,y,z$ is considered. The end-to-end vector correlation function is defined as
\begin{equation}
\phi(t)=3\left\langle \mathbf{R}^T_{ee,\alpha}(t)\mathbf{R}_{ee,\alpha}(0)\right\rangle,
\end{equation}
where the factor 3 accounts for the three-dimensional space. Using Eq. \eqref{eq: LangevinSolution} and \eqref{eq: Etebead}, the latter expression can be rewritten as
\begin{equation}\label{eq: PhiBeads}
\phi(t)= 3\left\langle  \mathbf{R}_{\alpha}^T(t)M\mathbf{R}_{\alpha}(0)\right\rangle= \left\langle \mathbf{R}_\alpha(0)^T e^{-\frac{At}{\tau}} M\mathbf{R}_{\alpha}(0)\right\rangle,
\end{equation}
where
\begin{equation}
M=\mathbf{v}\cdot \mathbf{v}^T=\begin{pmatrix}
1 & 0 & \cdots & 0 & -1\\
0 & 0 &  &  & 0\\
\vdots &  & \ddots &  & \vdots\\
0 &  &  & 0 & 0\\
-1 & 0 & \cdots & 0 & 1\\
\end{pmatrix}.
\end{equation}	
Note that the stochastic term appearing in Eq. \eqref{eq: LangevinSolution} vanishes when computing the average. Diagonalizing the matrix $ A $ and introducing our nondimensional units, we obtain
\begin{equation}\label{eq: Phisol}
\phi(t)=3\left\langle \mathbf{R}_\alpha(0)^T P^Te^{-{Dt}}P M\mathbf{R}_{\alpha}(0)\right\rangle.
\end{equation}
In contrast to the MSD, the end-to-end relaxation cannot be computed analytically, so we are forced to generate the bead positions of a large ensemble of chains in equilibrium (See SI for further details about how to do this in an efficient way).

\begin{figure}[ht]
	\centering
	\includegraphics[width=\columnwidth]{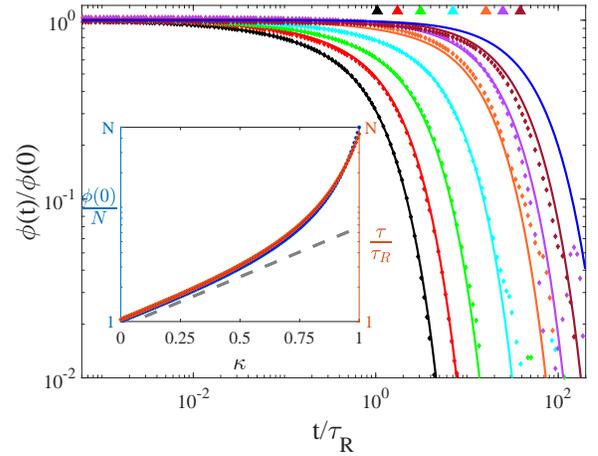}
	\caption{End-to-end relaxation function $ \phi(t) $ from BD simulations (symbols) and analytical calculations (solid lines) for $ N=50 $ and the same values of $\kappa$  displayed in the previous figure (the same color code is used). The terminal relaxation times of all cases are indicated in the top of the plot with triangles. The solid blue line represents the end-to-end relaxation of a rod polymer with constant length. The inset shows the static end-to-end distance divided by $N$ (blue) and relaxation time divided by $\tau_R$ (orange) as a function of the parameter $\kappa$. Dashed grey line depicts a perfect exponential scaling and is shown as a guide to the eye.}
	\label{fig: Fig3}
\end{figure}

Using simple geometrical arguments, we can obtain the initial value of the end-to-end vector relaxation, which, in terms of the bonding vectors, can be expressed as:
\begin{equation}\label{eq: Phi0Def}
\phi(0)=\left\langle\sum_{i=1}^{N}\mathbf{q}_i^2\right\rangle+2\left\langle\sum_{i=1}^{N}\sum_{j=1}^{i-1}\mathbf{q}_i\mathbf{q}_j\right\rangle.
\end{equation}
Using Eq.\eqref{eq: Productq}, the expression can be reduced and exactly calculated to give:
\begin{equation}\label{eq: Phi0}
\begin{split}
\phi(0)&=Nb^2+2b^2\sum_{i=1}^{N}\sum_{j=1}^{i-1}\kappa^{i-j}\\
& =Nb^2\left(1+2\kappa\frac{N(\kappa-1)+\kappa^N-1}{N(\kappa-1)^2}\right).
\end{split}
\end{equation}
Please note that the result is slightly different to that obtained by Winkler \textit{et al} for a similar model (see Eq. 3.48 in \cite{winkler1994}) using reduced constraints (see Eq. 3.22 and 3.23 in \cite{winkler1994}). 
In the rod-like limit ($\kappa\rightarrow1$) the end-to-end relaxation can be estimated using the rotational diffusion of a rod in dilution, which is the relaxation mechanism that allows the end-to-end vector to change its orientation. In chapter 8 of the book of Doi and Edwards\cite{doi1988theory}, an exact expression for the decay of the end-to-end vector of a rod polymer is provided. Neglecting hydrodynamic interactions, the rotational diffusion  is given by $ D_r = k_BT/\zeta_r $, where $ \zeta_r = \pi\eta_sL^3/4 $ and $L$ is the constant length of the rod. Hence, the end-to-end relaxation of a rod polymer, expressed in nondimensional units, is
\begin{equation}\label{eq:}
\phi(t)=N\exp\left(-\frac{8t}{N\pi^2}\right).
\end{equation}
In Fig. \ref{fig: Fig3}, we show the results from BD simulations as well as our analytical results, Eq. \eqref{eq: Phisol} for chains with $ N=50 $ beads. The time axis has been divided by $ \tau_R $ and $ \phi(t) $ by its equilibrium value $ \phi(0) $, taken from Eq. \eqref{eq: Phi0}.
The figure shows a perfect agreement between the numerical data from simulations and the analytical results from Eq. \eqref{eq: Phisol}. In contrast to the MSD of the central monomer, shown in Fig. \ref{fig: Fig2},  the growth of terminal relaxation time with the stiffness of the chain is more clear. In addition, the shape of $\phi(t)$ is almost single exponential regardless of value of the stiffness parameter $\kappa$ (it is exactly a single exponential for the purely rigid rod). Note that the semiflexible Gaussian chain in the stiff limit $\kappa\to 1$ is not equivalent to a rigid rod, because fluctuations in the chain contour length are permitted. The relaxation of the end-to-end vector is largely dependent of the magnitude of the end-to-end distance. As the chains get stiffer, the end-to-end vector grows and, thus, the terminal time increases. In the limit of $\kappa\to 1$, all beads lie on a straight line, but the contour length of the chain can fluctuate, whereas it is constant in the case of a pure rod. As a consequence, the end-to-end relaxation of a semiflexible Gaussian chain is faster than that of a rigid rod, because orientational diffusion is sped up when the length of the rod is shrunk by fluctuations. 

The growth of the equilibrium mean-square end-to-end distance with the rigidity parameter can be observed in the inset of Fig. \ref{fig: Fig3}. In the flexible limit ($\kappa=0$, Rouse chain), $ \phi(0)=\langle R^2\rangle = Nb^2$, whereas in the rod limit the function yields $ \phi(0)=N^2b^2 $. The funciton $\phi(0)$ grows exponentially with the stiffness parameter for $\kappa\lesssim 0.5 $ and even faster for stiffer chains. Thus, the effect of the rigidity is much more notable for values of $ \kappa $ close to 1. The same dependence can also be observed in the terminal relaxation time, although the pure rod relaxation time ($\tau=N\tau_R$) is never reached, due to the fluctuations of the contour length, as explained above.
%This singular behavior is explained by the vibrational modes present in our model, that gives a relaxation time smaller than $\tau$ of a pure continuous rod.

\subsection{Shear stress relaxation modulus}\label{subsec: Gt_semiflexible}

The shear stress relaxation modulus $ G(t) $ of the generalized semiflexible Gaussian chain can be calculated by using the  definition of $ G(t) $ from linear response theory:
\begin{equation}\label{eq: Gdef}
G(t) = \frac{V}{k_BT}\left\langle \sigma_{\alpha\beta}(t)\sigma_{\alpha\beta}(0)\right\rangle,
\end{equation}
where $\sigma_{\alpha\beta}$ is an off-diagonal component $(\alpha\beta)$ of the stress tensor and $ V $ is the volume. The contribution to the stress tensor of an individual chain is given by
\begin{equation}\label{eq: sigma}
	\sigma_{\alpha\beta}=\frac{1}{V}\mathbf{R}^T_\alpha\frac{\partial U}{\partial \mathbf{R}_\beta}=\frac{3k_BT}{Vb^2}\mathbf{R}^T_\alpha(t)A\mathbf{R}_\beta(t).
\end{equation}
Substituting Eq. \eqref{eq: sigma} in Eq. \eqref{eq: Gdef} and expanding the product with the explicit evolution of the beads position given by Eq. \eqref{eq: LangevinSolution}, we obtain a sum of four terms which can be simplified to (see SI)
\begin{equation}\label{eq: GDefNondim}
\begin{split}
	G^*(t)=\sum_{i,j}&\nu_i\nu_j e^{-2\nu_it}\left\langle x^{(i)}_\alpha x^{(j)}_\alpha x^{(i)}_\beta x^{(j)}_\beta\right\rangle     + \\  
	+ \;& \nu_j\delta_{\alpha\beta}\langle x^{(i)}_\alpha x^{(i)}_\beta\rangle\left(1-e^{-2\nu_it}\right),
\end{split}
\end{equation}
where $ x_\alpha^{(i)} $ is the $ i $th element of the normal mode $ \mathbf{X}_\alpha=P\mathbf{R}_\alpha $ and $G^*(t) = G(t) Vb^4/9k_B T$. Since we are interested in the shear stress, $ \alpha\neq \beta $, the second term in Eq. \eqref{eq: GDefNondim} is zero. In the expression above, the averages involving normal coordinates are obtained at equilibrium and
%, as in the case of the end-to-end relaxation function, they 
are the only terms in Eq. \eqref{eq: GDefNondim} that require a numerical computation. A similar expression as Eq. \eqref{eq: GDefNondim} is given in Ref. \cite{Dolgushev_2009}.

\begin{figure}[ht]
	\centering
	\includegraphics[width=\columnwidth]{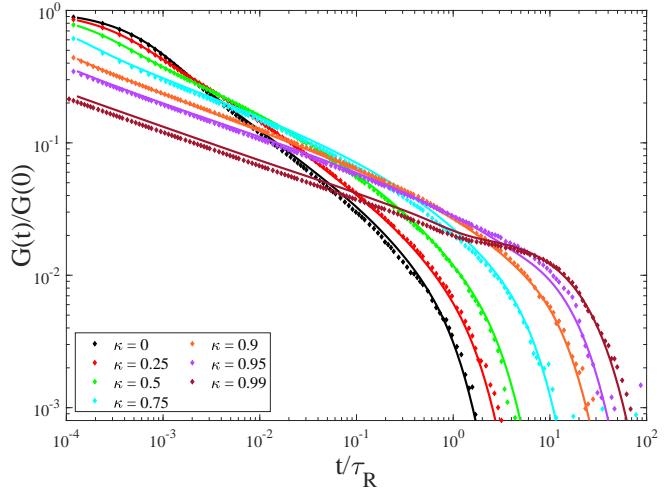}
	\caption{Shear stress relaxation modulus $G(t)/G(0) $ from BD simulations (symbols) and analytical calculations (solid lines) for $ N=50 $ and the same values of the stiffness parameter $\kappa$ displayed in previous figures.}
	\label{fig: Fig4}
\end{figure}

The normalized shear stress relaxation modulus from Eq. \eqref{eq: GDefNondim} and from BD simulations, shown in Fig. \ref{fig: Fig4}, show very good agreement except for very stiff chains, \textit{i.e.} $\kappa=0.99$, due to the tiny time step that must be used in the simulations, which limits the length of the simulations and their ability to explore the configurational space (See SI). This reduces the accuracy in the value of $G(0)$ and, in turn, of the accuracy of the complete relaxation modulus.

As in the MSD, increasing the stiffness has two effects on the shape of $G(t)$. On the one hand, it increases the amount of relaxed stress at short times. 
Please note that all curves in Fig. \ref{fig: Fig4} are normalized so that they start decaying from $1$. 
As the stiffness increases, the spectrum of relaxation times widens. Faster relaxation times are associated with bending modes, which become increasingly important as $\kappa$ gets larger. Thus, a large fraction of the initial stress is relaxed at a very early time. On the other hand, as the rigidity increases, the size of the chain grows and the terminal relaxation time gets longer. To mitigate its initial stress, a chain must completely forget its initial orientation and this occurs more slowly as the chain size increases. As a consequence of the initial, fast decay of $G(t)$ and the delay of the terminal decay, the characteristic power law $G(t)\propto t^{-1/2}$ of a Rouse chain gets softer for stiff chains and shows an approximate decay of $ G(t)\propto t^{-0.25} $. A similar slope is observed in the loss modulus for semiflexible chains with dihedral constraints \cite{Dolgushev_2013}. For very stiff chains, the relaxation spectra of $G(t)$ gets so wide that the relaxation modulus shows a small plateau before the terminal time, as can be observed for $\kappa=0.99$.

\section{\label{sec:whip}Whip-like polymer}

The method presented in the previous section is general enough so it can also be applied to other types of Gaussian chains, even if the values of the stiffness parameter $\kappa$ and the average bond length $b$ are not constant. In fact, if one can write the equations of motion for the bead positions as in Eq. \eqref{eq: LangevinMatrix}, the same expressions for all the observables derived in the previous section apply.
%, as well as the equilibrium conformations, of such system can be explicitly obtained. 
This approach does not only include polymers with arbitrary semiflexibility parameter $\kappa$ and bond length $b$ along the contour (see Supporting Information, section III), but also can embrace other systems subjected to other long-range correlations between bond vectors that may mimic torsions (see Supporting Information, section IV) or even active forces as long as they are linear with respect to the beads positions (see for instance Eq. (39) in our previous work\cite{tejedor2020}).
Here, we explore the behaviour of semiflexible Gaussian chains when the stiffness parameter $\kappa$ changes along the chain but the mean-square bond length $b^2$ is kept constant. A similar approach has been used to study the relaxation modulus of linear chains and dendrimers with arbitrary stiffness along their contour\cite{Dolgushev_2010}. Starting from the potential of the semiflexible model (Eq.\eqref{eq: PotSemiflexibleq}) 
%in terms of the bonding vectors $\mathbf{q}$
, we can enforce a variable stiffness along the contour by changing the value of the stiffness parameter in the constraints, i.e.
\begin{equation}\label{eq: WhipConstraint}
\begin{split}
\left\langle \mathbf{q}_i^2\right\rangle & =b^2,\quad i=1,...,N,\\
\left\langle \mathbf{q}_i\mathbf{q}_{i+1}\right\rangle & =b^2\kappa_i\quad 
i=1,...,N-1.
\end{split}
\end{equation}

As a result of changing the constraints, the Lagrangian multipliers $\lambda_i,\,\mu_i$ must be recalculated. 
%The constraints in Eq. \eqref{eq: WhipConstraint} differ from those in Eq. \eqref{eq: ConstrSemiflexible} and, therefore, the Lagrangian multipliers $\lambda_i,\,\mu_i$ must be recalculated. 
We compute the expected values using the maximum entropy principle and then we enforce the corresponding constraints. It is worth highlighting that the calculation does not depend on the precise values of the parameter $ \kappa_i $, as long as they take any value in the interval $ [-1,1] $. From the original paper of Winkler \cite{winkler1994}, we have an exact expression to compute any expectation value $\phi_k$ in the Hamiltonian of the system by using the partition function $Z$:
\begin{equation}\label{eq: WhipExp}
\phi_k=-\frac{\partial\log Z}{\partial \xi_k}=-\frac{1}{Z}\frac{\partial Z}{\partial 
\xi_k},
\end{equation}
where $\xi_k$ is the corresponding Lagrangian multiplier ($\lambda_i$ or $\mu_i$ in our system), and $\phi_k$ is the corresponding expectation value (whether $\langle\mathbf{q}_i^2\rangle$ or $\langle\mathbf{q}_i\mathbf{q}_{i+1}\rangle$). The partition function of our model in nondimensional units is defined as
\begin{equation}\label{eq: WhipZ}
Z=\int \exp\left( 
- \sum_{i=1}^{N}\lambda_i\mathbf{q}_i^2+\sum_{i=1}^{N-1}\mu_i\mathbf{q}_i\mathbf{q}_{i+1}\right)d^N\mathbf{q}.
\end{equation}
As in the case of the homogeneously semiflexible Gaussian chain, where the parameter $\kappa$ is constant along the contour, the exponential inside the integral of Eq. \eqref{eq: WhipZ} can be expressed in a bilinear form $\exp(-\mathbf{Q}^TB\mathbf{Q})$ with the semi-definite positive matrix $ B $, and the partition function can be computed as $Z=\pi^{N/2}|B|^{-1/2}$, where $|B|$ is the determinant of $B$.  For a given value of $ N $, the partition function and Eq. \eqref{eq: WhipExp} yield a system of nonlinear equations whose solutions are the Lagrangian multipliers which, in our case, take the form\cite{Dolgushev_2010}
\begin{equation}
\begin{split}
\lambda_i&=\frac{1}{2}\frac{1-k_{i-1}^2k_i^2}{(1-\kappa_{i-1}^2)(1-\kappa_i^2)},\quad 
i=2,...,N-1\\
\mu_i&=\frac{\kappa_i}{(1-\kappa_i^2)},\quad i=1,...,N-1,
\end{split}
\label{eq: LagMultipliersWhip}
\end{equation}
and $\lambda_1=\frac{1}{2(1-\kappa_1^2)}$, $\lambda_N=\frac{1}{2(1-\kappa_{N-1}^2)}$. Please note that the homogeneously semiflexible case is recovered by setting $\kappa_i=\kappa,\;\forall i$. It can be shown that these Lagrangian multipliers result in a semidefinite positive matrix $ B,\,\forall\kappa\in[-1,1]$.

\begin{figure}[ht]
	\centering
	\includegraphics[width=\columnwidth]{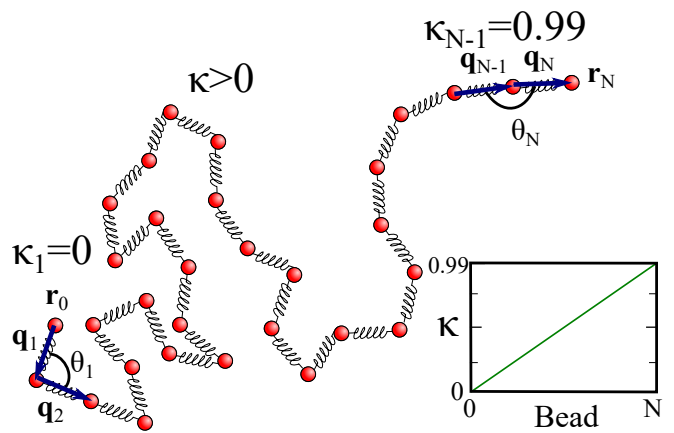}
	\caption{Schematic picture of the whip-like polymer indicating the totally flexible part $\langle \mathbf{q}_1\mathbf{q}_2\rangle_0=k_1=0$ and the rod-like extreme $\langle \mathbf{q}_{N-1}\mathbf{q}_{N}\rangle_0=k_{N-1}=0.99$. Inset depicts the value of $\kappa$ along the contour.}
	\label{fig: Fig6}
\end{figure}

We now explore the following particular case, depicted in Figure \ref{fig: Fig6}: the stiffness parameter grows linearly from one end, which is fully flexible ($ \langle\mathbf{q}_1\mathbf{q}_2\rangle=0 $), to the other end, which is almost completely stiff ($ \langle\mathbf{q}_{N-1}\mathbf{q}_{N}\rangle\approx 1 $). From now on, we will refer to this chain as the whip-like polymer, because the overall shape and flexibility of the chain reminds of a whip. Similar structures are commonly found in biological systems \cite{tournus2015, fauci1988, olson2011, montenegro2012, simons2015, schoeller2018}. For example, bacteria use cilia and flagella, which typically are linear structures with decreasing stiffness along their contours, to propel themselves through viscous media. Here, we are interested in studying the static and dynamical properties of whip-like polymers in dilute solution, neglecting hydrodynamic interactions and using the mathematical methods presented in the previous section, as well as by running BD simulations. 
As stated above, we set $\kappa_1=0$ (Rouse limit) and $\kappa_{N-1}=0.99$ (rod limit), and $\kappa_i=\kappa_1 + (i-1)(\kappa_{N-1}-\kappa_1)/(N-2)$. Please keep in mind that the rod-like limit $ \kappa=1 $ cannot be completely assessed with this formulation since the value of the Lagrangian multipliers diverge in that limit (see Eq. \eqref{eq: LagMultipliersWhip}. 

The whip-like chain can be treated similarly to the homogeneous semiflexible polymer, since the constant matrices $ A $ and $B$ that govern the dynamics keep the characteristics that facilitate its diagonalization, i.e. they are semidefinite-positive and symmetric, which holds as long as all $\kappa_i\in [-1,1]$. Thus, the expressions for the observables derived for the semiflexible model are still valid here: the MSD from Eq. \eqref{eq: MSDmFINsol}, the end-to-end autocorrelation function from  Eq. \eqref{eq: Phisol}, the shear stress relaxation modulus from Eq. \eqref{eq: GDefNondim} and the dynamic structure factor using Eqs. \eqref{eq: DSFdefSim} and \eqref{eq: DSFsol}.
In conclusion, the dynamical behavior is completely captured by the matrix of coefficients $A$, and this matrix encloses all geometrical constraints of the chain. In the following subsections, we explore the dynamics of whip-like polymers from both calculations and BD simulations, measuring the most typical observables and comparing them with those of homogeneously semiflexible Gaussian chains.

\subsection{Segmental motion}

The analytical solution for the MSD matrix $\mathbf{g}(t)$ can be exactly obtained from Eq. \eqref{eq: MSDmFINsol}, but the product $\left\langle\mathbf{R}\cdot\mathbf{R}^T\right\rangle_0$ requires a recalculation since Eq. \eqref{eq: ProductSemiflexible} does not hold for the whip-like polymer.  Similarly to the homogeneous semiflexible chain, the scalar product $ \langle\mathbf{r}_i\mathbf{r}_{j}\rangle_0 $ can be computed as a sum of products $ \langle\mathbf{q}_i\mathbf{q}_{j}\rangle_0 $. The result of Eq. \eqref{eq: Productq} can be generalized to the case of a semiflexible polymer with arbitrary stiffness parameters $\kappa_i$ along the contour as the product of all the stiffness parameters between bond \textit{i} and bond \textit{j}. Therefore, the scalar product between bonds is now expressed as
\begin{equation}\label{eq: ProductqWhip}
	\left\langle\mathbf{q}_i\mathbf{q}_j\right\rangle_0=\prod_{l=\min(i,j)}^{\max(i,j)-1}\kappa_{l}+\delta_{i,j}.
\end{equation}
Introducing this modification, it is again possible to obtain an analytical expression of the scalar product $\left\langle\mathbf{R}\mathbf{R}^T\right\rangle_0$. Hence, only the diagonalization of the constant matrix $ A $ requires numerical analysis.
\begin{figure}[ht]
	\centering
	\includegraphics[width=\columnwidth]{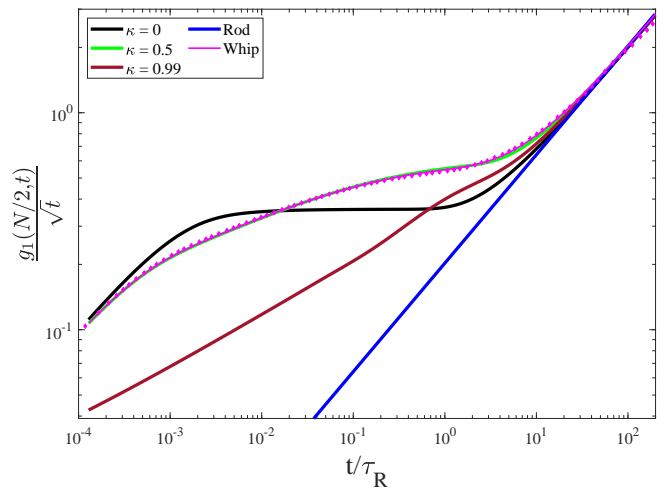}
	\caption{Mean-squared displacement of the middle bead for whip-like polymer (magenta) and semiflexible Gaussian chains with $ \kappa=0.0, 0.5, 0.99$, all with length $N=50$. Results are obtained from BD simulations (symbols) and from Eq. \eqref{eq: MSDmFINsol} (solid lines). The MSD is divided by $ \sqrt{t} $ and time is divided by $\tau_R$, the Rouse time of a flexible chain.}
	\label{fig: MSDWhip}
\end{figure}

Using this approach, the MSD of the middle bead of a whip-like polymer with $ N=50 $ is plotted in Fig. \ref{fig: MSDWhip}, compared to the MSD of the middle bead of semiflexible Gaussian chains with stiffness parameters $\kappa=$ 0, 0.5 and 0.99. 
Our results suggest that the MSD of the central monomer is almost identical to that of a homogeneously semiflexible polymer with $ \kappa=0.5 $, the mean value of the stiffness parameter in the whip-like chain. This result is reasonable because the local conformation and stiffness of the middle bead in both cases are almost equal and, thus, at early time, they must move in a similar way. However, at intermediate times, the dynamics of the middle bead is affected by the motion of all beads and, in principle, there is no reason why a whip-like chain should have the same MSD of the central monomer as the homogeneously semiflexible polymer. In fact, $g_1(N/2,t)$ in both models exhibits slightly different dynamics, as can be appreciated at intermediate times and at times near the Rouse time. It must be taken into account that the MSD measures the 2nd moment of the probability distribution of displacements. Probably, higher order moments of the pair-distribution function (PDF) must be inspected to observe any noticeable differences. Again, the diffusion of the center of mass does not depend on the value of the conformational stiffness along the chain. Therefore, in the Fickian regime, all segmental motions converge. 

\begin{figure}[ht]
	\centering
	\includegraphics[width=\columnwidth]{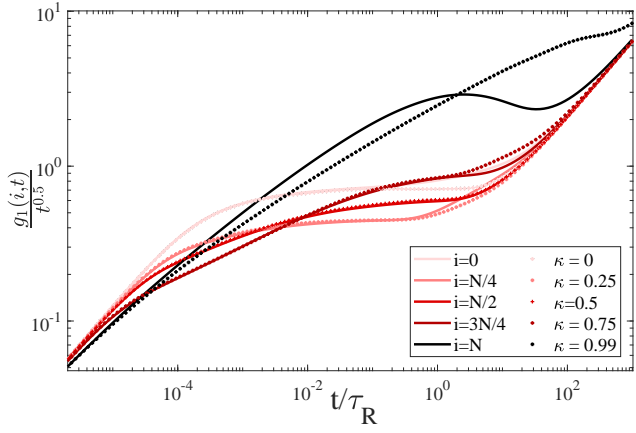}
	\caption{Mean-squared displacement (Eq. \eqref{eq: MSDmFINsol}) of different beads along a whip-like polymer with 200 beads (solid lines), and of the corresponding beads for homogeneously semiflexible polymers $ \kappa=0,0.25,0.5,0.75,0.99 $ (dotted lines). The MSD is divided by $\sqrt{t} $ and time is divided by $\tau_R $.}
	\label{fig: MSDWhip199}
\end{figure}

The head-tail asymmetry of the whip polymer, not reflected in the MSD of the middle bead, can be better appreciated by exploring the MSD of different segments along the whip polymer. The MSD of different beads along a chain with 200 beads are shown in Fig. \ref{fig: MSDWhip199}. To better highlight the heterogeneous dynamics of beads along the chain, we have considered a longer whip-like chain than that in Fig. \ref{fig: MSDWhip}. The MSD of each bead is compared with that of the analogous bead in a homogenously semiflexible polymer. For instance, the bead $ i=N/4 $ in the whip-like chain, with semiflexibility $ \kappa_{N/4}\approx0.25 $, is compared with the bead $ i=N/4 $ in a homogeneously semiflexible chain with $ \kappa=0.25 $.

As expected, our results reveal a clear head-tail asymmetric behavior with respect to the MSD of the center of the chain. Additionally, the MSD of beads in regions of the chain with low to medium rigidity are comparable to the MSD of the corresponding beads in a homogeneously semiflexible polymer, differing only in the region approaching the terminal time. More interestingly, in the terminal region, the MSD of flexible beads ($\kappa < 0.5$, light colors in Fig. \ref{fig: MSDWhip}) are faster than those of the corresponding bead, whereas in the case of beads with moderate rigidity ($\kappa=0.75$, dark red) the MSD is slower. On the other hand, at the rigid end of the whip-like polymer, the MSD is very different to that of the corresponding bead in a homogeneously semiflexible chain, being faster at early times and much slower close to the terminal region. To summarize, for beads that are closer to the middle bead, the agreement between the MSD with that of the corresponding homogeneously semiflexible bead is better. Notably, the MSD of the middle bead fits that of the middle bead of the homogeneous semiflexible chain, even better than for the shorter polymer ($ N=50 $) shown in Fig. \ref{fig: MSDWhip}. Therefore, we confirm that analyzing the MSD of the middle bead is not an appropriate observable to discern if a chain has constant or variable rigidity along its contour. 

Finally, it is noteworthy that the middle bead of the whip-like polymer is not the slowest moving monomer of the chain at all times, in contrast to the homogeneously semiflexible chain. For instance, at short times, beads at the stiff end are slower than the center, whereas, at times closer to the Rouse time, the slowest monomer in Fig. \ref{fig: MSDWhip199} is the bead $ i=N/4 $.

\subsection{Dynamic structure factor}

\begin{figure}[hbt!]
	\centering
	\includegraphics[width=\columnwidth]{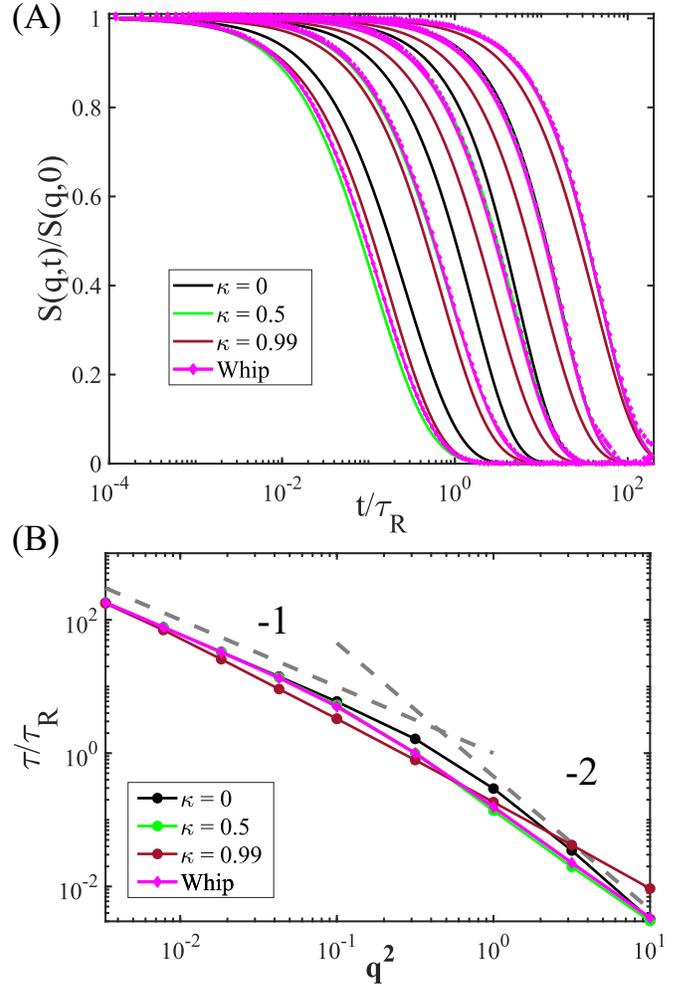}
	\caption{(A) Dynamic structure factor from BD simulations (pink symbols) and analytical calculations (pink solid lines) for the whip-like polymer with $ N=50 $ and homogeneously semiflexible chains with $\kappa=0,\,0.5,\,0.99$ using the same color code presented in previous figures. The scattering vectors are chosen to be $d^2=1,3,9, 27,81$, from left to right in the plot. (B) Relaxation time $\tau$ from the stretched exponential fit to $S(\mathbf{q},t)$ for three different values of $\kappa$ and the whip-like polymer with scattering vectors that span from long distance $d^2=300$ to length scales shorter than the bond distance $d^2=0.1$. Grey dotted lines show the power law of -1 and -2 as indicated.}
	\label{fig: Figure11}
\end{figure}

We can calculate the dynamic structure factor of the whip-like polymer and compare the results with different homogeneous cases. We contrast both results from simulations and from analytical calculations to the calculations of homogeneous semiflexible polymers with $\kappa=0,\,0.5$ and 0.99. In Fig. \ref{fig: Figure11} we plot the different curves with the same scattering vectors studied above: $d^2=1,\,3,\,9,\,27,\,81$.

Our analytical solution for $S(\mathbf{q},t)$ perfectly fits the simulation data for the whip-like polymer (see Fig. \ref{fig: Figure11}(A)). Both the homogeneous semiflexible and the whip-like polymer should converge to the Fickian dynamics of the center of mass for long enough scattering distances (much larger than the characteristic molecular size). Thus, we observe that the scattering vector $\mathbf{q}^2=1/81$ covers length scales greater than the corresponding static end-to-end distance and all curves agree, except for the stiffest chain with $\kappa=0.99$, which has a significantly larger size. 
Surprisingly, at intermediate distances, the whip-like polymer presents roughly the same dynamic structure factor as the homogeneous semiflexible chain with $\kappa=0.5$. This is unexpected, given the heterogeneous dynamics of the monomers along the whip-like polymer depicted in Fig. \ref{fig: MSDWhip199}. 
The results suggest that the section of the chain that is stiffer than the average $\kappa=0.5$ relaxes faster than the homogeneous case, whereas the more flexible section relaxes more slowly, and both sections compensate to yield the same curve of $S(\mathbf{q},t)$. 
Only at very short distances ($\mathbf{q}^2=1$), $S(\mathbf{q},t)$ of the whip-like polymers departs from that of the homogeneous counterpart. 
At those distances, the homogeneously flexible chain with $\kappa=0.5$ shows a minimum in the relaxation time (see Fig. \ref{fig: Figure5}B). 

We extract the characteristic relaxation time from the dynamic structure factor as a function of $\mathbf{q}$ using equation \eqref{eq: StretchedExponential}. As can be seen in Fig. \ref{fig: Figure11}(B)) the relaxation of the whip-like polymer is again very similar to the homogeneous semiflexible polymer with $\kappa=0.5$. Only at distances close to the bond length and below ($\mathbf{q}^2\gtrsim 1$) the relaxation parameter $\tau$ slightly differs from the homogeneous case. Overall, we observe that it is very difficult, by inspecting dynamical observables, to discern whether the stiffness of a polymer chain is homogeneous or not. Only by looking at the MSD of different monomers along the contour, it is possible to distinguish if the chain has stiffness heterogeneity. However, by definition, the equilibrium conformation is intrinsically distinct, so the end-to-end relaxation should reflect this underlying distinctness.

\subsection{End-to-end relaxation function}

\begin{figure}[ht]
	\centering
	\includegraphics[width=\columnwidth]{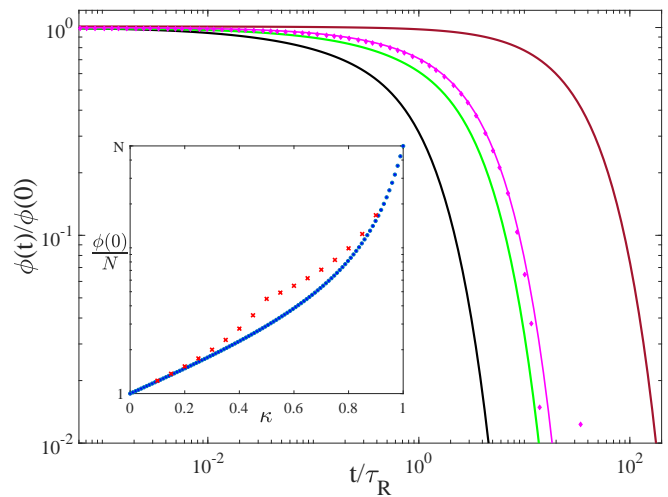}
	\caption{Log-log plot of the end-to-end correlation function $ \phi(t) /\phi(0) $ from BD simulations (symbols) and analytical calculations (solid lines) for whip-like and semiflexible chains with different values of $\kappa$ and $ N=50 $. The inset shows the equilibrium mean-square end-to-end distance (corresponding to $\phi(0)$) for homogeneously semiflexible (blue dots) and whip-like polymers of  average stiffness $\kappa$ (red symbols).}
	\label{fig: phiWhip}
\end{figure}

Similarly to the case of chains with homogeneous semiflexibility, here we analyze the end-to-end relaxation functions for linear chains with linearly varying stiffness along the backbone. We show the results from numeric calculation and BD simulations in Fig. \ref{fig: phiWhip}. To highlight the differences in the terminal time, the end-to-end relaxation function has been divided by the equilibrium mean-square end-to-end distance, and time has been divided by $\tau_R$. It is noteworthy that the shape of $\phi(t)$ for the whip-like polymer remains almost single exponential, in spite of the heterogeneous stiffness and more complex conformations of the chains. Compared to previously studied dynamical observables, the end-to-end vector of the whip-like polymer relaxes more slowly than the corresponding homogeneously semiflexible chain with $\kappa=0.5$. In order to clarify this, we inspect the dependence of the molecular size of the whip-polymer with respect to $\kappa$.

%Both the equilibrium mean-square end-to-end distance ($\phi(0)$) and the terminal time are slightly larger for the whip-like polymer. 

In the inset, the equilibrium values of the mean-square end-to-end distance are shown for chains with both homogeneous semiflexibility and whip-like polymers. In this plot, whip-like polymers with average semiflexibility $\kappa$ have been built by maximizing the span of the values of $\kappa_i$ along the contour, \textit{i.e.} selecting the limits $k_{min}$ and $k_{max}$:
\begin{equation}\label{eq: Heavy-Metal}
\begin{split}
\kappa_{min} &= 2 x H(x)  \\ 
\kappa_{max} &= 1 +  2x H(-x)
\end{split}
\end{equation}
with $x=(\langle \kappa \rangle - 0.5)$, $H(x)$ the Heaviside step function, and the values of $\kappa_i$ are interpolated linearly between $k_{min}$ and $k_{max}$ to get the desired $\langle \kappa \rangle$. 
%For example, if $\langle \kappa \rangle > 0.5$, we set $\kappa_{N-1}=0.99$ and determine $\kappa_1 = 2\langle\kappa\rangle -\kappa_{N-1}$, and if $\langle \kappa \rangle < 0.5$, we set $\kappa_1=0$ and determine $\kappa_{N-1} = 2\langle\kappa\rangle -\kappa_1$. 
Thus, we expect that when $\langle \kappa \rangle$ approaches either the Rouse or the rod limit, the behavior of the whip-like chain will be very similar to that of the corresponding chain with homogeneous stiffness.
%In this case we have divided $ \phi(t) $ by $ N $ to show the discrepancy between the whip polymer and the homogeneous semiflexible chain with $ \kappa=0.5 $.
The conformations of the whip-like polymer are more extended than the corresponding chains with homogeneous stiffness, and this effect is more appreciable when the average semiflexibility of the whip-like chain is around $ \kappa=0.5$. 
It is worth recalling that, for homogeneously semiflexible chains, the effect of stiffness on $\phi(0)$ and the terminal time increases more rapidly when $ \kappa $ approaches 1, (see Fig. \ref{fig: Fig3}). 
According to Eq. \eqref{eq: Heavy-Metal}, for intermediate values of $\langle\kappa\rangle$, the span of $\kappa$ values is maximized, and thus the global conformation and relaxation are mostly affected by the rigid section of the whip-like polymer. 
This suggests that the end-to-end vector relaxation, as well as the MSD of selected monomers, is a good indicator to discern if a chain has homogeneous or heterogeneous stiffness along the chain contour.

\subsection{Shear stress relaxation modulus}

\begin{figure}[ht]
	\centering
	\includegraphics[width=\columnwidth]{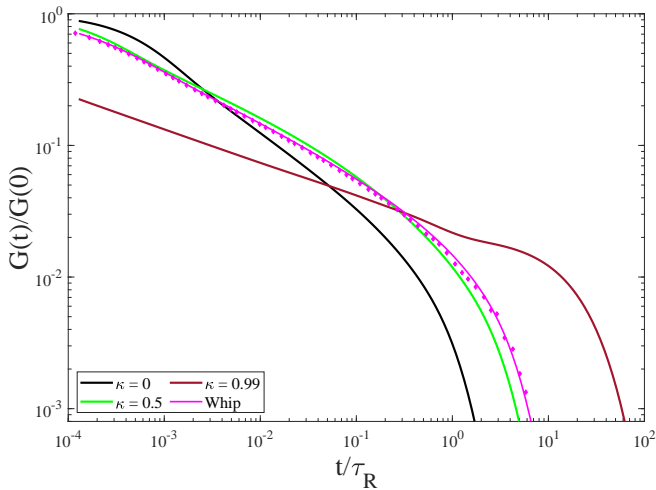}
	\caption{Log-log plot of the stress relaxation function $G(t)$ from BD simulations (symbols) and analytical calculations (solid lines) for $N=50$ and the same values of $\kappa$ displayed in the previous figure.}
	\label{fig: GtWhip}
\end{figure}

Finally, we study the shear stress relaxation function of the whip-like polymer, shown in Fig. \ref{fig: GtWhip}. Again, $G(t)$ of the whip-like chain is comparable to the semiflexible polymer with $ \kappa=0.5 $. However, if the curves are inspected in detail, they seem slightly different at intermediate as well as at long times. At short times, the contribution from chain bending to the relaxation modulus is larger in the whip-like than in the homogeneous chain and thus $G(t)$ decays faster given that, at that time scale, the stiffer section of the chain has a greater impact on $G(t)$ than the more flexible segments.  
However, in the terminal region, the final decay of $G(t)$ is governed by the relaxation of the end-to-end orientation. For this reason, the relaxation modulus of the whip-like polymer decays slightly slower than in the case of homogeneous stiff chains. Hence, the non-homogeneous flexibility induces small differences that can be appreciated in comparison with the homogeneous equivalent chain, but it is less sensitive than the MSD of selected beads.

\section{\label{sec:conclusion}Summary and Conclusions}

In this work, we have revisited a model of Gaussian semiflexible polymers that was proposed by Winkler\cite{winkler1994}, to derive some of the most important dynamical observables in the discrete limit. In particular, we provide analytical expressions for the relaxation modulus and the end-to-end vector relaxation (studied in previous works but with a different approach)\cite{Dolgushev_2009}, and for the segmental motion and dynamical structure factor. In addition, we run Brownian Dynamics simulations of the same systems to verify the accuracy of the proposed expressions. The bilinear form of the Hamiltonian allows us to generate initial configurations for the simulations exactly. 

Using this methodology, we first study polymers with homogeneous stiffness. We present a methodology to exactly compute different dynamical functions that perfectly match the behavior of such systems in Brownian dynamics simulations in all cases. We have verified through the MSD of the middle bead and the end-to-end vector relaxation that the system dynamics ranges from the Rouse-like to the rod-like limit, with the difference of a fluctuating contour in the limit of a rigid rod. More importantly, the dynamic structure factor provides a useful insight of the underlying conformation of homogeneous semiflexible chains, demonstrating a heterogeneous relaxation for all the range of distances depending on the stiffness of the chain. The fit of $S(\mathbf{q},t)$ to a stretched exponential reveals the expected dependence of the relaxation time with the scattering vector for flexible chains whereas a different tendency arise for the stiffest cases. Also, we show that the behavior of the chains is analogous when the explored distances are rescaled by the corresponding end-to-end distance and relaxation time.

%Parrafo homogeneous: 
%- MSD middle bead: A tiempos cortos, más lento cuanto más rígido. Cerca del tiempo terminal, se mueve más rápido que Rouse.
%- S(q,t): The relaxation time is not monotonous with respect to kappa at all q values. At long distances: the stiffer the faster. At short distances, slower at intermediate kappa. Gaussian relaxation at long distances (as expected, CM diffusion). Heterogeneous relaxation (stretched exponential) at intermediate distances for semiflexible chains. tau vs q: -2 at small q, -4 at long q. For very stiff chains this does not work. Much slower decay of tau. 
%At $q \ll R^2$, $tau \approx q^{-2}$. At $q \gg R^2$, $\tau \approx q^{-2.5}$ for very stiff.
%- End-to-end: Larger and slowed chains as the stiffness increases, much more evident the closer we get to 1.
%- G(t): Almost power law ($t^{-0.25}$) relaxation up until the terminal time. For very stiff chains, small plateau before exponential terminal decay. 

The expressions presented in the first part of the work are showed to be sufficiently general so they can be applied to any arbitrary model of polymer as far as its Hamiltonian can be expressed as a bilinear form. In that sense, we first generalize the system to a polymer with random stiffness, a chain with arbitrary stiffness and bond length, and a system with homogeneous stiffness and torsion-like bond correlations (the latter two cases are detailed in the Supplementary material). Then, we particularize all the dynamical equations to a particular case of a chain with variable stiffness that increases linearly from one end of the chain to the other (named as \textit{whip-like} polymer). 
 
We compare the dynamics of the whip-like polymer with that of homogeneously semiflexible chains, reporting a subtle difference with the case of homogeneous stiffness with $\kappa=0.5$. We show that most observables are equivalent to those of a homogeneous semiflexible chain with a constant semiflexibility parameter equal to the average $\kappa$ of the whip. However, the MSD of different beads along the contour, and the end-to-end relaxation help to discern the intrinsic asymmetry and heterogeneous dynamics of the whip-like polymer. 

%Observables with most differences: MSD, end-to-end. This may be good candidates to discern if a semiflexible polymer in solution has constant stiffness or not.

Overall, this works provides a straightforward methodology to study both the conformational and dynamical properties of discrete semiflexible Gaussian chains, which are a very simple but powerful model that can be compared with simulations of semiflexible molecules and can be used as a starting point for more complicated systems, such as entangled or charged polymers. We show that the same methodology can be applied to other cases of interest, provided that the Hamiltonian can be expressed as a bilinear form of the monomer coordinates. This includes longer distance correlations between the bonds, or different architectures such as branched polymers, rings or dendrimers \cite{Dolgushev_2009, Dolgushev_2010}. We believe that such family of models could be the most  appropriate to describe the dynamics of coarse-grained semiflexible chains, where bonds could have some degree of extensibility, whereas in the WLC bonds are inextensible.

%- Model to compare simulation results (simulation of semiflexible macromolecules).
%- Starting point for more complicated physics, such as entanglements or charged polymers. 
%The same methodology can be applied to other cases of interest, provided that the Hamiltonian can be expressed as a bilinear form of the monomer coordinates. This includes: longer distance correlations between the bonds (torsions?) or different architectures such as branched polymers, rings or dendrimers \cite{Dolgushev_2009, Dolgushev_2010}. 
%- It could be a more appropriate model to describe the dynamics of coarse-grained semiflexible chains, where bonds could have some degree of extensibility, whereas in the WLC bonds are inextensible. 

\section{Acknowledgements}
The authors acknowledge funding from the Spanish Ministry of Economy and Competitivity (PID2019-105898GA-C22) the computer resources and technical assistance provided by the Centro de Supercomputacion y Visualizacion de Madrid (CeSViMa).

\section{Data Availability}
The data that supports the findings of this study are available within the article and its supplementary material.

\section{References}
\bibliography{bibliography} 

\end{document}